\documentclass{article}

\usepackage[final]{neurips_2019}

\usepackage[utf8]{inputenc} 
\usepackage[T1]{fontenc}    
\usepackage{hyperref}       
\usepackage{url}            
\usepackage{booktabs}       
\usepackage{amsfonts}       
\usepackage{nicefrac}       
\usepackage{microtype}      
\usepackage{caption}

\usepackage{tabularx}
\usepackage{booktabs} 
\usepackage{array}
\usepackage{subfigure}
\usepackage{listings}
\usepackage{xcolor}
\usepackage{mathrsfs}
\usepackage{fancyhdr}
\usepackage{multirow}
\usepackage{enumitem}
\usepackage{authblk}
 \usepackage{lipsum}

\newcommand\blfootnote[1]{%
  \begingroup
  \renewcommand\thefootnote{}\footnote{#1}%
  \addtocounter{footnote}{-1}%
  \endgroup
}
 
\usepackage{titlesec}
\titlespacing\section{0pt}{12pt plus 4pt minus 2pt}{0pt plus 2pt minus 2pt}

\titlespacing\subsection{0pt}{4pt plus 0pt minus 4pt}{0pt plus 2pt minus 2pt}

\usepackage[norndcorners,customcolors]{hf-tikz}
\hfsetbordercolor{yellow}
\hfsetfillcolor{yellow}
\definecolor{mygreen}{rgb}{0,0.6,0}
\definecolor{mygray}{rgb}{0.5,0.5,0.5}
\definecolor{mymauve}{rgb}{0.58,0,0.82}

\title{Algorithm-hardware Co-design for Deformable Convolution}

\author{
\textbf{Qijing Huang*, \ Dequan Wang*, \ Yizhao Gao\textsuperscript{1}, \ Yaohui Cai\textsuperscript{2}, \ Zhen Dong, \ Bichen Wu, 
\ \ \ Kurt Keutzer, \ John Wawrzynek}\\
  University of California, Berkeley,\\ \textsuperscript{1}University of Chinese Academy of Science, \textsuperscript{2}Peking University\\
  \texttt{\{qijing.huang, dequanwang, zhendong, bichen, keutzer, johnw\}@berkeley.edu} \\
  \texttt{gaoyizhao16@mails.ucas.ac.cn, caiyaohui@pku.edu.cn}\\

}

\begin{document}
\blfootnote{EMC$^2$: 5th Edition Co-located with NeurIPS’19}
\maketitle
\vspace{-0.7cm}
\begin{abstract}

FPGAs provide a flexible and efficient platform to accelerate rapidly-changing algorithms for computer vision. The majority of existing work focuses on accelerating image classification, while other fundamental vision problems, including object detection and instance segmentation, have not been adequately addressed. Compared with image classification, detection problems are more sensitive to the spatial variance of objects, and therefore, require specialized convolutions to aggregate spatial information. To address this, recent work proposes dynamic deformable convolution to augment regular convolutions. Regular convolutions process a fixed grid of pixels across all the spatial locations in an image, while dynamic deformable convolutions may access arbitrary pixels in the image and the access pattern is input-dependent and varies per spatial location. These properties lead to inefficient memory accesses of inputs with existing hardware.
In this work, we first investigate the overhead of the deformable 
convolution on embedded FPGA SoCs, and then show the
accuracy-latency tradeoffs for a set of algorithm 
modifications including full versus depthwise, fixed-shape, and limited-range. 
These modifications benefit the energy efficiency for embedded devices in general as they reduce the compute complexity.  
We then build an efficient object detection network with modified deformable convolutions and quantize the network using state-of-the-art quantization methods.
We implement a unified hardware engine on FPGA to support all the operations in the network. 
Preliminary experiments show that little accuracy is compromised and speedup can be achieved with our co-design optimization for the deformable convolution.


\end{abstract}
\vspace{-0.5cm}
\section{Introduction}

Deep neural networks have achieved a series of successes in object recognition tasks, such as image classification
, semantic segmentation
, and object detection
.
Convolution, as its basic operation, is widely adopted in different neural network architecture designs. 
Many hardware accelerators 
have been developed to improve the speed and power performance of the compute-intensive convolutional kernels. 
While the use of convolution kernels for computer vision is well-established,
researchers have been constantly proposing new operations and new network designs
to increase the model capacity and achieve higher accuracy for various tasks.
Some of these new operations lack mature and efficient support from existing hardware.
FPGA, as a programmable platform, lends itself to accelerating fast-evolving deep learning algorithms. Timely and efficient hardware support for novel operations can be developed on FPGAs in weeks with high-level design tools. 

Deformable convolution~\cite{dai2017deformable, zhu2019deformable} is a novel and effective operation leading to the state-of-the-art accuracy for object recognition tasks. 
Three out of five implementations leading to top accuracy to date for object detection on the COCO dataset use deformable convolution in their design, including the 1st-ranked model ~\cite{chen2019hybrid}. Differing from the conventional convolutions with fixed geometric structure, deformable convolution samples inputs from variable offsets generated based on the input features of the target task. There are two advantages it provides to improve its modeling capacity compared to conventional convolutions: \textit{variable sampling scales} and \textit{variable sampling geometry}. The range for sampling at each different point varies, allowing the network to capture objects of different scales. The geometry of the sample points is not fixed, allowing the network to capture objects of different shapes. 
A previous study \cite{zhou2019objects} has also shown that the deformable convolution network (DCN) design lies on the pareto front of the speed-accuracy tradeoff for object detection on GPUs.

There are several challenges in supporting deformable convolution on hardware accelerators. 1) there is an increased memory bandwidth requirement for loading the variable offsets. 
2) the memory accesses for the input feature maps are irregular as they depend on the dynamically-generated offsets.
Many existing accelerators' instruction set architecture and the control logic are insufficient in supporting the random memory access patterns. 
In addition, the less contiguous memory access patterns limit the length of bursting memory accesses and incur more memory requests.   
3) there is less spatial reuse for the input features. 
Due to the variable filter offsets, the loaded input pixel for the current output pixel can no longer be reused by its neighboring output pixels. 
This can significantly affect the performance of the accelerators designed for output-stationary or row-stationary dataflow which leverages input reuse. 

To address these challenges, we adopt an algorithm-hardware co-design approach and study the accuracy-efficiency tradeoffs for each algorithmic modification.  
We propose the following modifications to the deformable convolution operation to make it more hardware friendly:

\begin{enumerate}[leftmargin=1cm]
    \item Remove the bilinear interpolation required for calculating the final sampling address
    \item Use only a square shape filter with variable dilation to reduce the total overhead from loading the offsets and to enable parallel accesses to on-chip memory
    \item Limit the dynamic sampling offsets to be positive and below a fixed value to allow buffering of inputs and exploit full input reuse. 
    \item Convert to depth-wise convolution to reduce total FLOPs
\end{enumerate}

We evaluate each modification on an FPGA System-on-Chip (SoC) that includes both an FPGA fabric and a hardened CPU core. We leverage the shared last-level cache (LLC) included in its full hardened processor system to efficiently exploit the locality from the original deformable convolution with data-dependent memory access patterns. We then implement each algorithm modification in hardware to demonstrate its effectiveness compared to the original design. With these proposed algorithm modifications, we devise a line-buffer design to efficiently support the modified depthwise deformable convolution. 
To demonstrate the full capability of the co-designed operation, we are in the process of training an efficient neural network model for object detection using  ShuffleNetV2 as the feature extractor.
We plan to quantize the network with a symmetric uniform quantizer using the block-wise quantization-aware fine-tuning process~\cite{dong2019hawq}.
Our main goals include: 
\begin{enumerate}[leftmargin=1cm]
    \item Co-design and evaluate a new depthwise deformable convolution with hardware-friendly modifications 
    \item Optimize the hardware design for each algorithm modification and evaluate the performance in accelerator systems with and without an LLC.  
    \item Design an efficient neural network model with the proposed depthwise deformable convolution for object detection, and quantize it to ultra-low previsions. 
    \item Implement a hardware accelerator targeting the new network design on an FPGA SoC.
\end{enumerate}
Promising results are shown from a preliminary study of algorithm-hardware co-design for deformable convolution in this paper.

\section{Deformable Convolution Co-design}
\label{deform_conv}
It is challenging to provide efficient support for the 
original deformable convolution on hardware due to: 
1) the increased computation for the bilinear interpolation, 
2) the overhead of the deformable offsets, 3) the dynamic and irregular input access patterns, 4) limited reuse of the input features. We perform a series of modifications to the algorithm to make it more hardware-friendly. For each modification, an experiment targeting the semantic segmentation task is run to test its impact on accuracy. 
We target semantic segmentation in this study as it shares the same feature extraction design but takes less time to train the end-to-end network compared to object detection. 
More importantly, the model capability to capture the geometric variation also affects the accuracy of semantic segmentation. 
We then design a specialized hardware engine optimized for each algorithmic modification on FPGA and show the performance improvement on FPGA from each modification. 
We demonstrate the accuracy and hardware efficiency trade-off for each modification we propose. 
\subsection{Algorithm Modifications}
Table~\ref{tab:algo_codesign} lists all the modifications we make to the deformable convolution and the corresponding mean Intersection over Union (mIoU) for the semantic segmentation task on the Cityscapes dataset~\cite{cordts2016cityscapes}. The Deep Layer Aggregation (DLA)~\cite{yu2018deep} network is used as the feature extractor in these experiments. 

\begin{figure}[!t]
\begin{minipage}[t]{0.20\linewidth}
\centering
	\includegraphics[width=\linewidth]{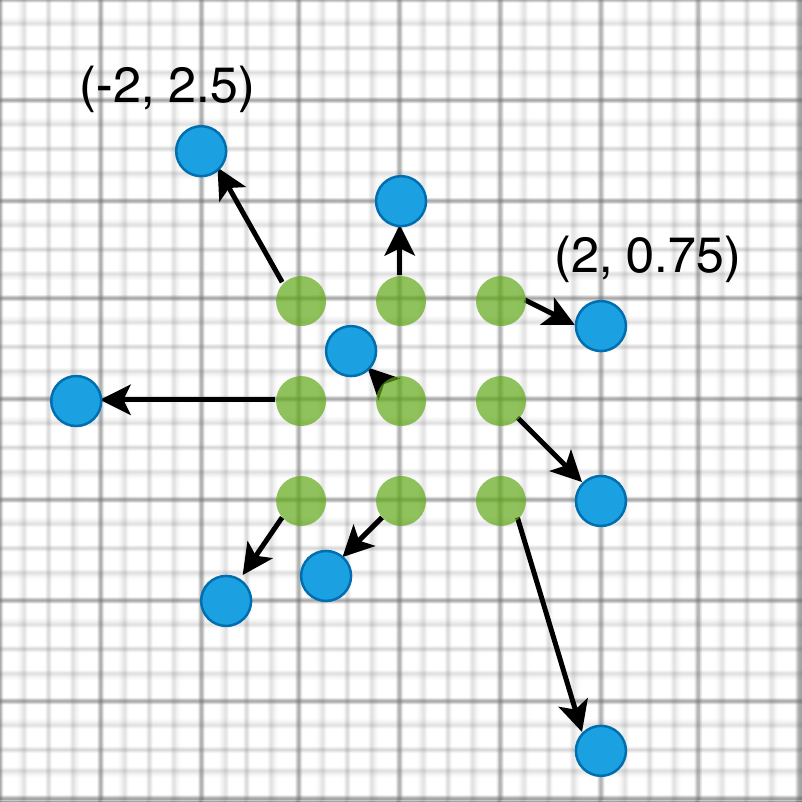}
	\vspace{-0.6cm}
	\caption*{a. Deformable}
\end{minipage}
\hspace{0.7cm}
\begin{minipage}[t]{0.20\linewidth} 
	\centering
	\includegraphics[width=\linewidth]{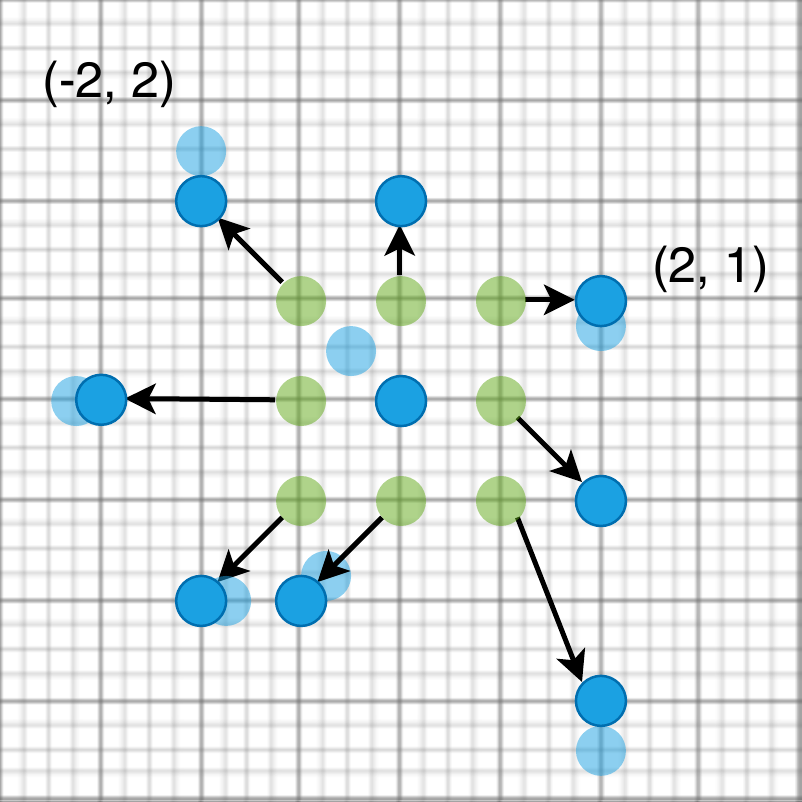} 
	\vspace{-0.6cm}
	\caption*{b. Round}
\end{minipage}   
\hspace{0.7cm}
\begin{minipage}[t]{0.20\linewidth}
\centering
	\includegraphics[width=\linewidth]{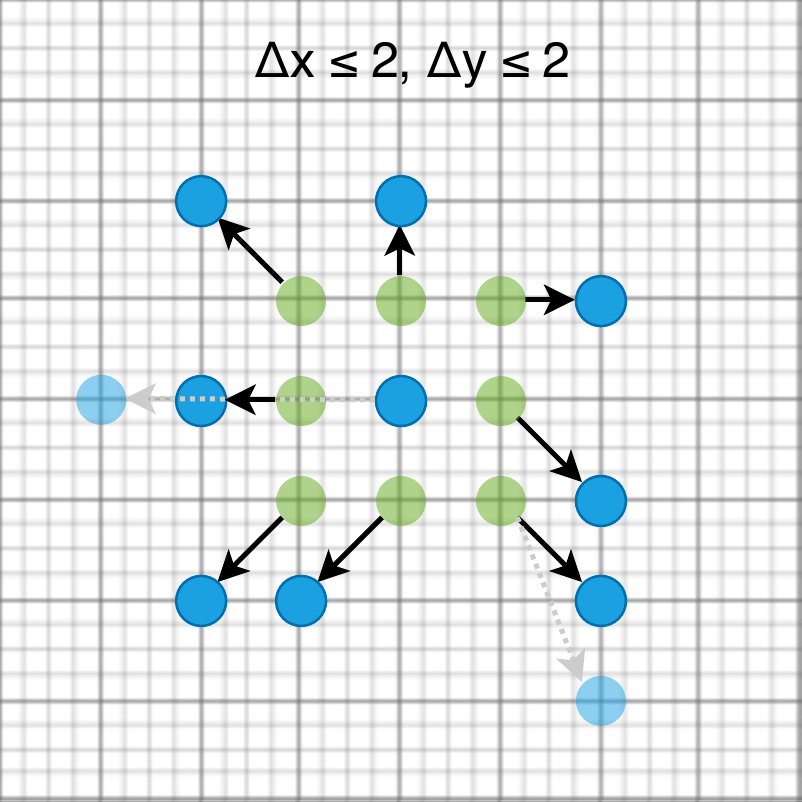}
	\vspace{-0.6cm}
	\caption*{c. Bound}
\end{minipage}
\hspace{0.7cm}
\begin{minipage}[t]{0.20\linewidth} 
	\centering
	\includegraphics[width=\linewidth]{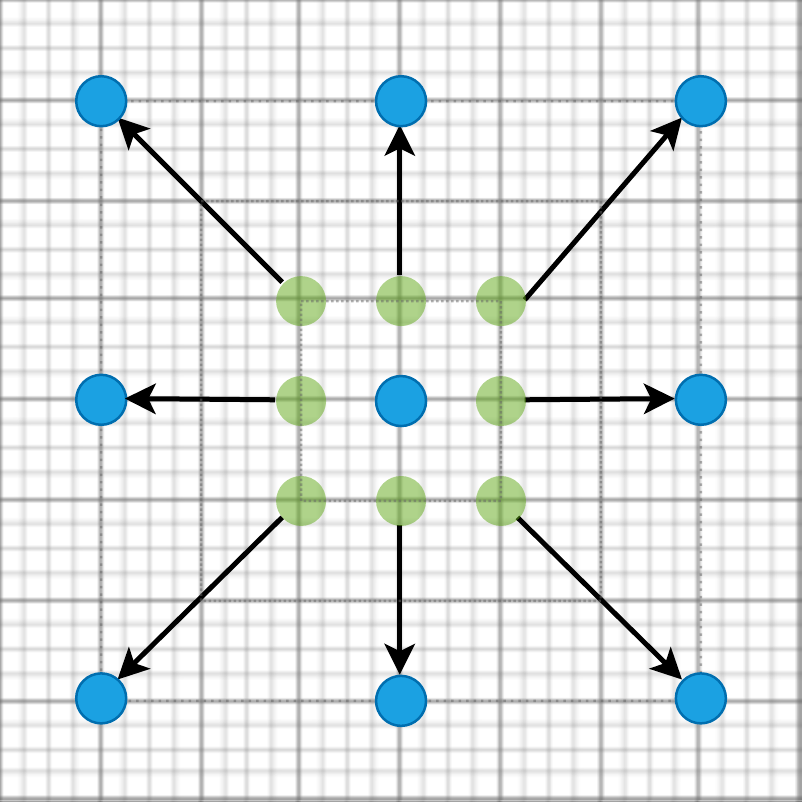} 
	\vspace{-0.6cm}
	\caption*{d. Square}	
 \end{minipage}  
\caption{Major Algorithm Modifications. a) is the original deformable convolution with non-integer offsets. b) shows that the offsets are rounded to integers. c) sets an upper bound to the offsets. d) limits the geometry to a square shape.}
\label{fig:algo_change}
\end{figure}  
 
\begin{table} 
\centering
\vspace{-0.6cm}
\caption{Notations}
\label{tab:notations}
\begin{tabular}{|c|c|c|c|c|c|c|}
\hline
$n$ & $h$ & $w$ & $ic$ & $oc$ & $k$ & $\Delta p$ \\ \hline
batch size & height & width & input channels size & output channel size & kernel size & offsets \\ \hline
\end{tabular}
\vspace{-0.7cm}
\end{table}

\begin{table} 
\vspace{-0.7cm}
\caption*{\textbf{Deformable Network Accuracy for Semantic Segmentation on Cityscapes Dataset}}
\hspace*{-1.8cm}
\begin{minipage}{.5\linewidth}
\centering
\caption{with DLA as Feature Extractor}
\label{tab:algo_codesign}
\begin{tabular}{|c|c|c|c|c|}
\hline
\textbf{Deformable} & \textbf{Round} & \textbf{Bound} & \textbf{Square} & \textbf{mIoU $\uparrow$} \\\hline
\checkmark & & & & 79.9 \\
\checkmark & \checkmark & & & 79.6 \\
\checkmark & \checkmark & \checkmark & & 79.4 \\
\checkmark & \checkmark & \checkmark & \checkmark & 78.7 \\
\hline
\end{tabular}
\end{minipage}%
\hspace*{1.8cm}
\begin{minipage}{.45\linewidth}
\centering
\caption{with Different Feature Extractors}
\label{tab:algo_codesign2}
\begin{tabular}{|c|c|c|}
\hline
\textbf{Feature Extractor} & \textbf{Operation} & \textbf{mIoU $\uparrow$} \\\hline
DLA & DeformConv & 79.9 \\
ShuffleNetV2 & DeformConv & 70.1 \\
ShuffleNetV2 & DeformConv + Depthwise & 68.0 \\
\hline
\end{tabular}
\end{minipage}%
\end{table}

\begin{table} 

\end{table}
 
\textbf{Rounded Offsets}. In the original deformable design as shown in Figure~\ref{fig:algo_change}a, the generated offsets are typically fractional and a bilinear interpolation needs to be performed to produce the final sample value from the input. 
Bilinear interpolation calculates a weighted average of the neighboring pixels for a fractional offset based on its distance to the neighboring pixels. 
It introduces at least six multiplications to the sampling process of each input, which is a significant increase ($6 \times h \times w \times ic $) to the total FLOPs. We thus round the offsets to be integers during inference to reduce the total computation. 
The dynamically-generated offsets are thus rounded to integers as illustrated in Figure~\ref{fig:algo_change}b. 
As shown in Table~\ref{tab:algo_codesign}, there is only a 0.3 drop in mIoU with this change.   

\textbf{Bounded Range}.
Another algorithmic modification to facilitate efficient hardware acceleration is to restrict the offsets to a positive range as shown in Figure~\ref{fig:algo_change}c. 
This limits the size of the working set of inputs, so a fixed-size buffer can be added to the hardware to further exploit the temporal and spatial locality of the inputs. 
Assume a uniform distribution for the generated offsets in a $3\times3$ convolution kernel with stride $1$, each pixel is expected to be used nine times. 
If all inputs within the range can be stored in the buffer, all except the first accesses to the same address will be from on-chip memory with $1 \sim 3$ cycle latency. 
We impose this constraint during training by adding a \textit{ReLU6} layer after the offset generation layer to truncate offsets that are smaller than 0 or larger than $N$, so all offsets $\Delta p_x, \Delta p_y \in [0, N]$.
Setting the bound $N$ to 7 results in an insignificant 0.2 mIoU degradation.

\textbf{Square Shape}. 
One obstacle to efficiently supporting the deformable convolution is its irregular data access patterns, which leads to serialized memory accesses to multi-banked on-chip memory. To address this issue, we further constrain the offsets to be on the edges of a square. Instead of using 18 numbers to represent the $\Delta p_x$ and $\Delta p_y$ offsets for all nine samples, only one number $\Delta p_d$,   representing the distance from the center to the sides of the square, needs to be learned. This is similar to a dilated convolution with variable dilation factors for each output. 
This modification leads to a 0.6 decrease in mIoU. Together with all the modifications above, we achieve an mIoU of 78.7, which is only 1.2 lower compared to the original implementation.  

\textbf{Feature Extractor Selection}. 
In order to further reduce the total FLOPs of the network to suit our real-time requirement, we change the feature extractor from DLA to ShuffleNetV2~\cite{ma2018shufflenet}. This results in a 9.8 decrease in mIoU, but almost $\sim$7$\times$ reduction in FLOPs (1.06 vs 0.146 GFLOPs). 

\textbf{Depthwise Convolution}.
We further replace all the full 3$\times$3 deformable convolutions with 3$\times$3 depthwise deformable convolutions and 1$\times$1 convolutions similar to the ShuffleNetV2 building block design. This change leads to another 2.1 mIoU drop as shown in Table~\ref{tab:algo_codesign2}. Using the ShuffleNetV2-like structure to replace the original deformable convolutions also makes the network more uniform and easier to support on FPGAs.

\begin{table}[!t]
\vspace{-0.5cm}
\centering
\caption{Co-designed Hardware Performance Comparison}
\hspace*{-1.8cm}
\begin{tabular}{|c|c|c|c|c|c|c|c|c|}
    \hline
\multirow{2}{*}{ \textbf{Operation}}& \textbf{Without} & \multirow{2}{*}{ \textbf{Deformable}} &  \textbf{Bound } &  \textbf{Square} &
\multicolumn{2}{c|}{ \textbf{Without LLC}} & \multicolumn{2}{c|}{\textbf{With LLC}} \\
\cline{6-9}
& \textbf{Deformable} & & (buffered) & (multi-ported) & Latency (ms) & GOPs & Latency (ms) & GOPs \\
\hline
& \checkmark & & & & 43.1 & 112.0 & 41.6 & 116.2 \\
Full&  & \checkmark & & & 59.0 & 81.8 & 42.7 & 113.1 \\
3$\times$3 Conv &  & \checkmark & \checkmark & & 43.4 & 111.5 & 41.8 & 115.5 \\
&  & \checkmark & \checkmark & \checkmark & 43.4 & 111.5 & 41.8 & 115.6 \\
\hline
& \checkmark & & & & 1.9 & 9.7 & 2.0 & 9.6 \\
Depthwise&  & \checkmark & & & 20.5 & 0.9 & 17.8 & 1.1 \\
3$\times$3 Conv&  & \checkmark & \checkmark & & 3.0 & 6.2 & 3.4 & 5.5 \\
&  & \checkmark & \checkmark & \checkmark & 2.1 & 9.2 & 2.3 & 8.2 \\
\hline
\end{tabular}
\label{tab:hw_codesign}
\end{table}

\begin{figure}[!t]
\vspace{-0.4cm}
\centering
	\includegraphics[width=0.6\linewidth]{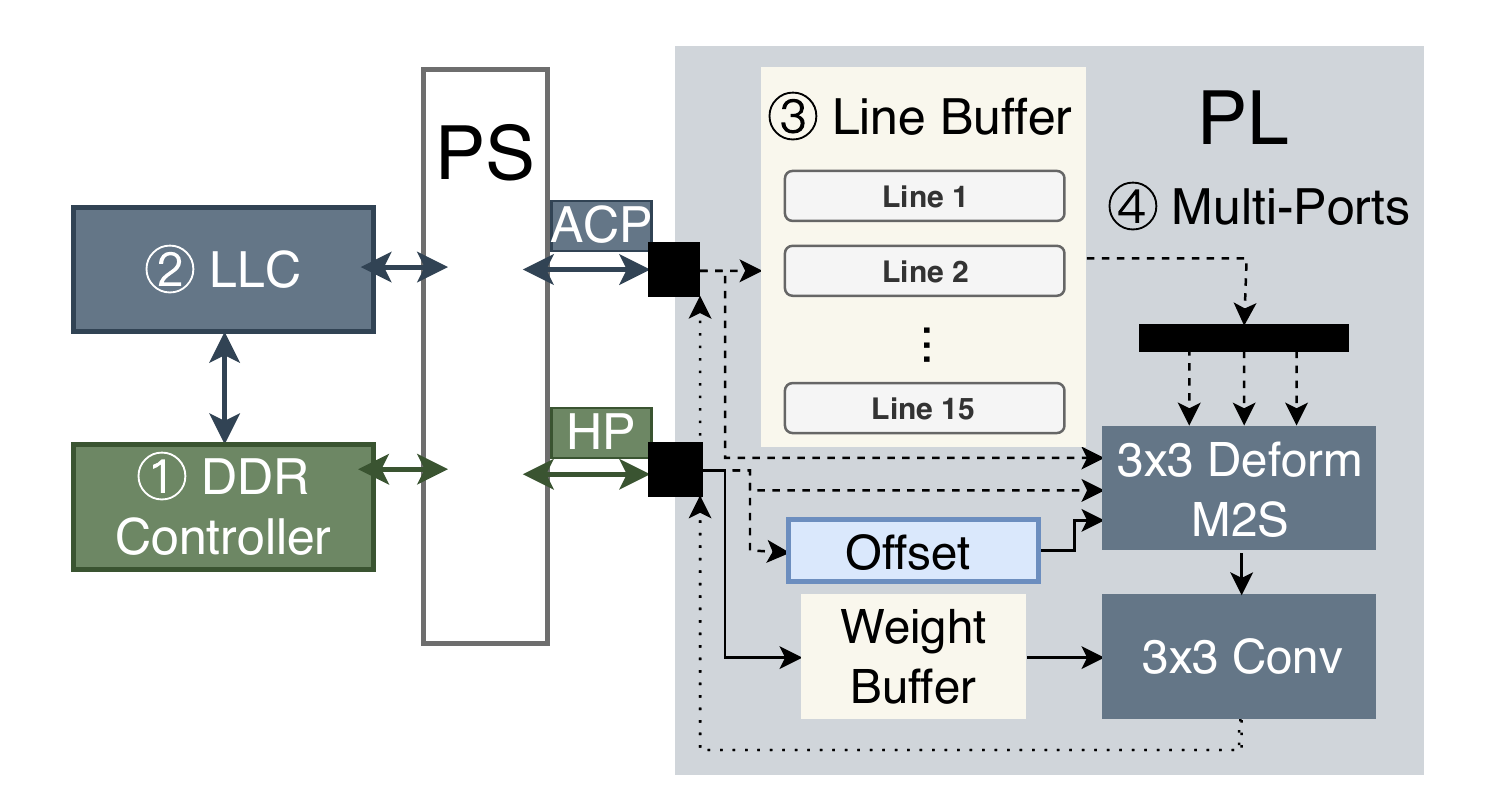}
\caption{Hardare Engine for Deformable Convolution}
\label{fig:hw_codesign}
\vspace{-0.7cm}
\end{figure} 
 
\subsection{Hardware Optimizations}
Many hardware optimization opportunities are exposed after we perform the aforementioned modifications to deformable convolution. We implement a hardware deformable convolution engine on FPGA SoC as shown in Figure~\ref{fig:hw_codesign} and tailor the hardware engine to each algorithm modification. The experiments are run on the Ultra96 board featuring a Xilinx Zynq XCZU3EG UltraScale+ MPSoC platform. The accelerator logic accesses the 1MB 16-way set-associative LLC through the Accelerator Coherency Port (ACP).  The data cache uses a pseudo-random replacement policy.  
Table~\ref{tab:hw_codesign} lists the speed and throughput performance for different customized hardware running a kernel of size $h=64, w=64, k=256, c=256$. 
In all experiments, we round the dynamically-generated offsets to integers. We use $8\times8\times9$ Multiply-Accumulate (MAC) units in the $3\times3$ convolution engine for all full convolution experiments and $16\times9$ MACs for depthwise convolution experiments.

\textbf{Baseline}.
The baseline hardware implementation for the original $3\times3$ deformable convolution directly access the DRAM wihtout going through any cache or buffering. In Figure 2, the baseline implementation directly access the input and output data through HP ports and  \textcircled{1} DDR controller.
The input addresses are first calculated from the offsets loaded from DRAM. The \textit{$3\times3$ Deform M2S} engine then fetches and packs the inputs into parallel data streams to feed into the the MAC units in the \textit{$3\times3$ Conv} engine. 


\textbf{Caching}. 
One hardware optimization to leverage the temporal and spatial locality of the nonuniform input accesses is to add a cache to the accelerator system. 
As shown in Figure~\ref{fig:hw_codesign}, we load the inputs from \textcircled{2} LLC through the ACP port in this implementation to reduce the memory access latency of the cached values.  
Since the inputs are sampled from offsets without specific patterns in the original deformable convolution, the cache provides adequate support to buffer inputs that might be reused in the near future. 
As shown in Table~\ref{tab:hw_codesign}, adding LLC results in 26.7\% and 13.2\% reduction in latency for the original full and depthwise deformable convolution respectively. 

\textbf{Buffering}. 
With the bounded range modification to the algorithm, we are able to use the on-chip memory to buffer all possible inputs. 
Similar to a line-buffer design for the original $3\times3$ convolution that stores two lines of inputs 
to exploit all input locality, we store $2N$ lines of inputs so that it is sufficient to buffer all possible inputs for reuse. This implementation includes the \textcircled{3} Line Buffer in Figure~\ref{fig:hw_codesign}.
With the effective buffering strategy, we can see in Table~\ref{tab:hw_codesign} that the latency of a bounded deformable is reduced by 26.4\% and 87.5\% for full and depthwise convolution respectively in a system without LLC. In a system with LLC, the reduction is 2.1\% and 80.9\% respectively.  The depthwise deformable convolution benefits more from adding the buffer as it is a more memory-bound operation. The compute-to-comunication ratio for its input is $oc$ times lower than the full convolution. 

\textbf{Parallel Ports}
The algorithm change to enforce a square-shape sampling pattern not only reduces the bandwidth requirements for loading the input indices in hardware, but also helps to improve the on-chip memory bandwidth.  
With non-predictable memory access pattern to the on-chip memory, only one input can be loaded from the buffer at each cycle if all sampled inputs are store in the same line buffer. 
By constraining the shape of deformable convolution to a square with variable dilation,
we are guaranteed to have three different line buffers with each storing three sampled points.
We can thus have three parallel ports (\textcircled{4} Multi-ports in Figure~\ref{fig:hw_codesign}) accessing different line buffers concurrently. 
This co-optimization improves the on-chip memory bandwidth and leads to another $\sim$ 30\% reduction in latency for depthwise deformable convolution. 


\section{Conclusion}
In this work, we perform a detailed accuracy-efficiency trade-off study for each hardware-friendly algorithmic modification to the deformable convolution operation, with the goal of co-designing an efficient object recognition network and a real-time embedded accelerator optimizing for accuracy, speed, and energy efficiency.
Experimental results show that these modifications lead to significant hardware performance improvement with little accuracy loss. 

\bibliographystyle{unsrtnat}
\bibliography{ref}
\end{document}